# Reconstructed standard model of cosmology in the Earth-related coordinate system

Jian-Miin Liu
Department of Physics, Nanjing University
Nanjing, The People's Republic of China
On leave. E-mail: liu@phys.uri.edu

**Abstract**

In the Earth-related coordinate system, we reconstruct the standard model of cosmology based on the assumption of the cosmological principle and the perfect gas (or fluid). We exactly solve Einstein's field equation involved. The exact solution consists of three parts respectively on the line element for space-time of the Universe, the value for the cosmological constant and the equation of state for the matter of the Universe.

1. **Introduction**

The current standard model of cosmology starts with the Friedmann-Robertson-Walker line element of the four-dimensional space-time,

$$ds^2 = c^2 dt^2 - R^2(t)\{\frac{dr^2}{1-kr^2} + r^2 d\theta^2 + r^2 \sin^2\theta d\phi^2\}, \qquad (1)$$

under the assumption of the large-scale homogeneity and isotropy of space [1-12]. This line element is a one having been represented in the co-moving coordinate system which, at any point in space, is at rest with respect to the matter located at that point [9,10]. Just in the co-moving coordinate system, the time-time component of the Friedmann-Robertson-Walker metric is assigned to be time-independent, $g_{00}=1$. However, this line element is not a good starting point.

Measuring and describing motion is always relative to a certain frame of reference. We have a lot of different frames of reference to select for the purpose, various inertial frames of reference and various non-inertial frames of reference. Moreover, an observer binding to the selected frame of reference may establish different coordinate systems to interpret his measurements and descriptions, like Cartesian coordinate system and spherical coordinate system. These different coordinate systems of the selected frame of reference contain their own deterministic contents [13,14]. What is the co-moving coordinate system? It is a coordinate system formed by connecting the small pieces of coordinate system that belong to different frames of reference, which each is at rest with respect to the matter located in its own small piece. How do we perform experimental measurements? During our experimental measurements, do we change the frame of reference, from time to time, making it at rest with respect to the measured objects? No, we never do so. An observer or a group of observers binding to one frame of reference usually performs experimental measurements. The co-moving coordinate system cannot be an established coordinate system of the one frame of reference in which experimental measurements are performed. In the framework of Einstein's theory of gravitation, due to its equivalence principle and general covariance, we can select any frame of reference and adopt any coordinate system in theoretical calculations, including the co-moving coordinate system. But, a question will arise when the co-moving coordinate system is adopted. How do we make comparison between calculation results and measurement data? We need, in principle, to know the transformation between the co-moving coordinate system and the established coordinate system of the one frame of reference. Once we know this transformation, we can transform all calculation results from the co-moving coordinate system to the established coordinate system of the one frame of reference and compare them to measurement data there. The said transformation is a necessary step. The current standard model of cosmology seems to miss it.

In the present paper, we do not intend to make this step for the current standard model of cosmology. We prefer to start again. The paper consists of five sections: introduction, Earth-related frame of reference, cosmological principle and perfect gas, Einstein's equation in the gravitational field of the Universe and its exact solution, concluding remarks.

2. **Earth-inertial frame of reference**

We are living on the Earth. We call such a frame of reference, in which the Earth is always at rest, the Earth-related frame of reference. In the presence of any gravitational field, the Earth-related frame of reference cannot be an inertial frame of reference. But, when all gravitational field is turned off, the Earth-related frame of reference is at least approximately an inertial frame of reference.

We have selected the Earth-related frame of reference for the purpose of cosmological measurements. We now select it for the purpose of theoretical calculations and descriptions.

3. **Cosmological principle and perfect gas**

To be able to apply Einstein's theory of gravitation to the study of the universe as a whole, we need to simplify the physical system of the Universe. We are going on the assumption of the cosmological principle. It says: With respect to the Earth-related frame of reference, the universe is spatially homogeneous and isotropic on large scales. Here, the homogeneity and isotropy are of the notion of the average over large region of space.

A theorem [5] states that, for an N-dimensional space having its maximally symmetric subspace of M dimensions, we are able to choose a coordinate system consisting of M $u^p$-coordinates, $p=1,2,---,M$, and N-M $v^a$-coordinates, $a=1,2,---,$N-M, so that the line element of the N-dimensional space has the form,

$$ds^2 = g_{ab}(v)dv^a dv^b + f(v)g_{pq}(u)du^p du^q, \qquad (2)$$

where $g_{ab}(v)$ and $f(v)$ are both functions of only $v^a$-coordinates, $g_{pq}(u)$ is the metric of the M-dimensional subspace which reflects the maximally symmetric properties of the subspace.

The Universe existing in the four-dimensional space-time with signature (+,-,-,-) has the three-dimensional space of the large-scale homogeneity and isotropy with respect to the Earth-related frame of reference. According to the theorem, there must be an Earth-related coordinate system, denoted by $\{x^i\}$, $i=0,1,2,3$, $x^0=ct$, in which the line element of the four-dimensional space-time of the Universe can be written as

$$ds^2 = S^2(t)c^2 dt^2 - R^2(t)g_{rs}(x)dx^r dx^s, \quad r,s=1,2,3,$$

where $c$ is the speed of light in vacuum, $S^2(t)$ is taken for $g_{ab}(v)$ while $R^2(t)$ for $f(v)$, and line element $g_{rs}(x)dx^r dx^s$ will reflect the large-scale homogeneity and isotropy of space. Evidently, the large-scale homogeneity and isotropy of space, or the symmetric properties under the spherical and displacement transformations in space, confine line element $g_{rs}(x)dx^r dx^s$ to be

$$\frac{dr^2}{1-kr^2} + r^2 d\theta^2 + r^2 \sin^2\theta d\phi^2,$$

in the spatially-spherical coordinates $x^1 = r$, $x^2 = \theta$, $x^3 = \phi$, where $k$ is the spatial curvature which takes value of +1 or 0 or -1, separately characterizing a closed or flat or open space. In this way, the line element of space-time of the Universe obeying the cosmological principle is determined up to two arbitrary time-dependent functions, $S^2(t)$ and $R^2(t)$, i.e.

$$ds^2 = S^2(t)c^2 dt^2 - R^2(t)\{\frac{dr^2}{1-kr^2} + r^2 d\theta^2 + r^2 \sin^2\theta d\phi^2\}, \quad (3)$$
$$S^2(t) > 0, \ R^2(t) > 0,$$

in the Earth-related coordinate system $\{ct, r, \theta, \phi\}$, where $S^2(t) > 0$ and $R^2(t) > 0$ are for keeping the signature. We abandon the co-moving coordinate system. Instead, we take the Earth-related coordinate system at the expense of one more time-dependent factor to be determined.

The cosmological principle also implies that the Universe is filled with the gravity-interacting matter which distributes uniformly in space. That means that non-zero mass density ρ of the matter of the Universe is independent of spatial coordinates. It is independent of time coordinate, too. Otherwise, we will face the problem that mass would be created (or annihilated) with the same rate everywhere in an arbitrarily big but finite spatial region.

The matter of the Universe is also treated as a perfect gas (or fluid): Its velocity vanishes everywhere with respect to the Earth-related frame of reference. This treatment is of course a simplification but reasonable because the most probable velocity in the perfect gas is small due to low cosmic background temperature, 2.7K. It can be in fact regarded as an outcome of the average over large region of space. As a result, the energy-momentum tensor for the matter of the Universe is simply

$$T^\mu_\nu = \begin{cases} \rho, & for \mu = \nu = 0, \\ -p\delta^\mu_\nu, & for \mu = \nu = 1,2,3, \\ 0, & otherwise, \end{cases} \quad (4)$$

everywhere in the Earth-related coordinate system, where $p$ is the pressure of the perfect gas. We take the system of units of $c = 1$ here and after unless specified. Such a kind of the matter of the Universe produces a gravitational field that we call the gravitational field of the Universe. The gravitational field of the Universe is also the notion of the average over large region of space.

4. **Einstein's equation in the gravitational field of the Universe and its exact solution**

We take the Einstein field equation with a cosmological term [15,16],

$$R^\mu_\nu - \frac{1}{2} R^\lambda_\lambda \delta^\mu_\nu = 8\pi G T^\mu_\nu + \Lambda \delta^\mu_\nu, \quad \mu = \nu = 0,1,2,3, \quad (5)$$

where G is the Newtonian gravitational constant, $\Lambda$ is the cosmological constant, $R^\mu_\nu$ is the Ricci curvature tensor and $R^\lambda_\lambda$ is the space-time curvature. Calling line element Eq.(3), we compute:

$$\Gamma^0_{00} = \frac{\dot{S}}{S}, \qquad \Gamma^1_{01} = \Gamma^2_{02} = \Gamma^3_{03} = \frac{\dot{R}}{R},$$

$$\Gamma^0_{11} = \frac{R\dot{R}}{S^2(1-kr^2)}, \qquad \Gamma^0_{22} = \frac{r^2 R\dot{R}}{S^2}, \qquad \Gamma^0_{33} = \frac{r^2 \sin^2\theta R\dot{R}}{S^2},$$

$$\Gamma^1_{11}=\frac{kr}{1-kr^2}, \qquad \Gamma^1_{22}=-r(1-r^2), \qquad \Gamma^1_{33}=-r(1-kr^2)\sin^2\theta,$$

$$\Gamma^2_{12}=\Gamma^3_{13}=\frac{1}{r}, \qquad \Gamma^2_{33}=-\frac{1}{2}\sin 2\theta, \qquad \Gamma^3_{23}=\cot\theta,$$

and other components vanish, where all "$(t)$" of $S(t)$ and $R(t)$ have been dropped for simplicity and the dot refers to a derivative with respect to time. From these expressions, we can further compute the Ricci tensor and the space-time curvature. Their non-vanishing components are

$$R^0_0 = \frac{3}{S^2 R}(-\ddot{R}+\frac{\dot{S}\dot{R}}{S}), \tag{6a}$$

$$R^1_1 = R^2_2 = R^3_3 = -\frac{2\dot{R}\dot{R}}{R^2 S^2} - \frac{R\ddot{R}}{R^2 S^2} - \frac{2k}{R^2} + \frac{\dot{R}\dot{S}}{RS^3}, \tag{6b}$$

and

$$R^\lambda_\lambda = \frac{6}{S^2 R}(-\ddot{R}-\frac{\dot{R}}{R}-\frac{kS^2}{R}+\frac{\dot{S}\dot{R}}{S}). \tag{7}$$

We put Eqs.(4), (6a-b) and (7) into Eq.(5) and obtain

$$\dot{R}^2 + k\,S^2 = \frac{8}{3}\pi\rho G\,R^2\,S^2 + \frac{1}{3}\Lambda R^2\,S^2, \tag{8a}$$

$$2S\,R\,\ddot{R} + S\,\dot{R}^2 + k\,S^3 - 2R\,\dot{R}\,\dot{S} = -8\pi\,p\,G S^3\,R^2 + \Lambda S^3\,R^2. \tag{8b}$$

as the dynamical equations of the Universe. The set of these equations reduces to that of the Friedmann equations if you recognize $S=1$ [1-12].

Multiplying Eq.(8a) by $S$ and inserting the resultant equation in Eq.(8b), we have

$$S\,R\,\ddot{R} - R\,\dot{R}\,\dot{S} = -\frac{4}{3}\pi\rho G\,S^3\,R^2 - 4\pi\,p\,G S^3\,R^2 + \frac{1}{3}\Lambda S^3\,R^2. \tag{9}$$

As said, the Earth-related coordinate system is at least approximately an inertial coordinate system when the gravitational field of the Universe is turned off. So, if the Universe was empty, the Einstein field equation would give us the Minkowskian metric as its solution. Setting $\rho=p=0$ and $S=R=1$ in Eq.(9), we find

$$\Lambda=0. \tag{10}$$

Once again, setting $\rho=0$ and $S=R=1$ in Eq.(8a) and using Eq.(10), we get

$$k=0. \tag{11}$$

The dynamical equations of the Universe then become

$$S = (\frac{3}{8\pi\rho G})^{1/2} \frac{\dot{R}}{R}, \tag{12a}$$

$$2 S R \ddot{R} + S \dot{R}^2 - 2 R \dot{R} \dot{S} = -8\pi p G S^3 R^2. \tag{12b}$$

Using Eq.(12a) in Eq.(12b) immediately yields

$$3(\frac{3}{8\pi\rho G})^{1/2} \frac{\dot{R}^3}{R}(1+\frac{p}{\rho})=0. \tag{13}$$

Either $\dot{R}$ or $\rho + p$ must be equal to zero. If $\dot{R}=0$, $S$ equals zero in accordance with Eq.(12a). That fails to meet the requirement of $S^2 > 0$. The only choice is therefore

$$\rho + p = 0. \tag{14}$$

It is so far so good but only one equation is left for us to determine two time-dependent factors $R$ and $S$. We have to secure an appropriate mathematical or physical constraint which we can combine with the left equation for determination of $R$ and $S$.

In the line element Eq.(3), $ds$ is an invariant four-dimensional "distance". In the case of no gravitational field, according to special relativity, this $ds$ shapes in

$$ds^2 = c^2 dt^2 (1 - \delta_{rs} \frac{dx^r}{cdt} \frac{dx^s}{cdt}). \tag{15}$$

It characterizes the proper time in the definition that the proper time of an event is a time interval as measured by the clock of an observer who is at rest relative to the event. The proper time is equal to the coordinate time $dt$ when spatial increments $dx^r$ or velocity $dx^r/dt$, $r=1,2,3$, vanish. If the concept of $ds$ characterizing the proper time is also applicable to the case of gravitational field, we have such appropriate constraint. As a perfect gas, the matter of the Universe has vanishing velocity everywhere in the Earth-related coordinate system. This gives out

$$S = 1. \tag{16}$$

We insert Eq.(16) in Eq.(12a) and find its unique solution

$$R = \exp[\sqrt{\frac{8\pi\rho G}{3}}(t - t_0)], \tag{17}$$

where $t_0$ is a constant. Putting Eqs.(11), (16) and (17) into Eq.(3), we obtain the line element of the Universe,

$$ds^2 = c^2 dt^2 - \exp[2\sqrt{\frac{8\pi\rho G}{3}}(t - t_0)]\{dr^2 + r^2 d\theta^2 + r^2 \sin^2\theta d\phi^2\}, \tag{18}$$

in the Earth-related coordinate system.

In view of Eq.(18), there was or will be a moment, at which the line element of the Universe was or will be the Minkowskian. That moment is $t_0$. We call it the SC-moment. We are entitled to choose the origin of the Earth-related time coordinate provided we do not change its scale. We choose the SC-moment as the origin of the Earth-related time coordinate. In doing so, in the Earth-related coordinate system, the line element of the Universe reads

$$ds^2 = c^2 dt^2 - \exp[2\sqrt{\frac{8\pi\rho G}{3}}t]\{dr^2 + r^2 d\theta^2 + r^2 \sin^2\theta d\phi^2\}, \qquad (19)$$

where $t$ is relative to the SC-moment.

Eqs.(10), (14) and (19) form the exact solution to Einstein's equation in the gravitational field of the Universe. Eq.(10) specifies the cosmological constant to be zero. Eq.(14) exhibits the equation of state for the perfect gas of the Universe. Eq.(19) is the line element for space-time of the Universe.

## 5. Concluding remarks

We have reconstructed, in the Earth-related coordinate system, the standard model of cosmology. We have exactly solved Einstein's field equation involved. From this exact solution we draw some conclusions as follows.

(1) In the gravitational field of the Universe, the line element of space-time of the Universe varies with time. It is in time-evolution. In this evolution, there was or will be the SC-moment, at which the line element was or will be the Minkowskian. Time is running forever from the past to the future. We have to depend on experimental observations to determine whether the SC-moment was in the past or will be in the future.

(2) The line element of space-time of the Universe stands at all times, from $-\infty$ to $+\infty$. We can see no need of introducing the concept of the age of the Universe.

(3) Eqs.(11), (16) and (17) allow us to calculate the space-time curvature of the Universe in Eq.(7),

$$R_\lambda^\lambda = -16\pi\rho G - 6\sqrt{8\pi\rho G/3}\exp(-\sqrt{8\pi\rho G/3}t).$$

It varies with time, too. It is negative for always and never becomes zero unless $\rho = 0$. It approaches $-16\pi\rho G$ eventually. The four-dimensional space-time of the Universe is curved forever though its three-dimensional space is flat.

(4) The equation of state, Eq.(14), indicates that the matter of the Universe, as a perfect gas, possesses a negative pressure.

(5) It seems that to add the cosmological term to the original version [15] of Einstein's field equation is not necessary when we apply Einstein's theory of gravitation to the study of the gravitational field of the Universe, as well as, to the study of the spherically symmetric gravitational field [17].

The obtained exact solution is effective in the studies of only those physical processes that occur in large region of space because Eqs.(3) and (4) lose their validity in any localized small region of space. A physical process occurring in a localized small region of space must be under control of some gravitational field other than the gravitational field of the Universe or some interaction other than gravity. We shall use the obtained exact solution to derive the formula relating the red-shift of light signals coming from distant galaxies to the distance of these galaxies from us and the time of detecting of these light signals. This formula will help us in understanding the observed "accelerating expansion of the Universe". We report these somewhere else [18].

**Acknowledgment**

The author greatly appreciates the teachings of Prof. Wo-Te Shen. The author thanks to Dr. J. Conway for his supports and helps.**References**

[1]  A. Friedmann, Z. Phys., 10, 377 (1922)
[2]  G. Lemaitre, Ann. Soc. Sci. (Bruxelles), 47, 49 (1927)
[3]  H. P. Robertson, Appl. Phys. J., 82, 284 (1935); 83, 187 (1936); 83, 257 (1936)
[4]  A. G. Walker, Proc. Lond. Math. Soc., 42, 90 (1936)
[5]  S. Weinberg, Gravitation and Cosmology, Wiley & Sons (New York, 1972)
[6]  E. W. Kolb and M. S. Turner, The Early Universe, Addison-Wesley (Redwood City, CA, 1990)
[7]  B. S. Ryden, Introduction to Cosmology, Addison-Wesley (Redwood City, 2002)
[8]  S. Dodelson, Modern Cosmology, Academic Press (New York, 2003)
[9]  S. K. Bose, An Introduction to General Relativity, Wiley & Sons (New York, 1980)
[10] M. Trodden and S. M. Carroll, astro-ph/0401547
[11] J. Lesgourgues, astro-ph/0409426
[12] J. Garcia-Bellido, astro-ph/0502139
[13] A. Einstein, Autobiographical Notes, in: A. Einstein: Philospheo-Scientist, ed. P. A. Schipp, 3rd edition, Tudor, New York (1970)
[14] Jian-Miin Liu, physics/0208047
[15] A. Einstein, Ann. der Phys., 49, 769 (1916)
[16] A. Einstein, Sitzungsber. Preuss. Akad. Wiss. 235 (1931)
[17] Jian-Miin Liu, A Test of Einstein's Theory of Gravitation: Equilibrium Velocity Distribution of Low-Energy Particles in Spherically Symmetric Gravitational Field, in Fronties in Field Theory, ed. O. Kovras, Nova Science Publishers, Inc. (Hauppauge, NY, 2005) [gr-qc/0206047, 0405048]
[18] Jian-Miin Liu, Formula for red-shift of light signals coming from distant galaxies, to be published [astro-ph/0505xxx]7